\documentclass[aps,prb,twocolumn,superscriptaddress]{revtex4}
\usepackage{dcolumn} % Align table columns on decimal point
\usepackage{graphicx}% Include figure files
\usepackage{bm}      % bold math
\usepackage{color}
\usepackage{amsmath,amssymb}
\usepackage{hyperref}
\usepackage{multirow}
\usepackage{epstopdf}
\usepackage{latexsym}
\usepackage{subfigure}
\usepackage{wasysym} %for hexagon (in Eq environment or in text)
\usepackage{times}
\usepackage{gensymb}
\usepackage{textcomp}

\begin{document}
\title{Magnetoelectric effect arising from a field-induced pseudo Jahn-Teller distortion in a rare earth magnet}
\author{Minseong Lee}
\email{minseong.lee10k@gmail.com}
\affiliation{Department of Physics, Florida State University, Tallahassee, Florida 32306-3016, USA}
\affiliation{National High Magnetic Field Laboratory, Florida State University, Tallahassee, Florida 32310-3706, USA}
\affiliation{National High Magnetic Field Laboratory, Los Alamos National Laboratory, Los Alamos, NM 87545, USA}

\author{Q. Chen}
\affiliation{Department of Physics and Astronomy, University of Tennessee, Knoxville, Tennessee 37996, USA}

\author{Eun Sang Choi}
\affiliation{National High Magnetic Field Laboratory, Florida State University, Tallahassee, Florida 32310-3706, USA}

\author{Q. Huang}
\affiliation{Department of Physics and Astronomy, University of Tennessee, Knoxville, Tennessee 37996, USA}

\author{Zhe Wang}
\affiliation{Anhui Key Laboratory of Condensed Matter Physics at Extreme Conditions, High Magnetic Field Laboratory, Chinese Academy of Sciences, Hefei, Anhui 230031, China}
\affiliation{University of Science and Technology of China, Hefei, Anhui 230026, China}

\author{Langsheng Ling}
\affiliation{Anhui Key Laboratory of Condensed Matter Physics at Extreme Conditions, High Magnetic Field Laboratory, Chinese Academy of Sciences, Hefei, Anhui 230031, China}

\author{Zhe Qu}
\affiliation{Anhui Key Laboratory of Condensed Matter Physics at Extreme Conditions, High Magnetic Field Laboratory, Chinese Academy of Sciences, Hefei, Anhui 230031, China}
\affiliation{University of Science and Technology of China, Hefei, Anhui 230026, China}

\author{G. H. Wang}
\affiliation{Department of Physics and Astronomy, Shanghai Jiao Tong University, Shanghai 200240, China}

\author{J. Ma}
\affiliation{Key Laboratory of Artificial Structures and Quantum Control, School of Physics and Astronomy, Shanghai Jiao Tong University, Shanghai 200240, China}

\author{A.A. Aczel}
\email{aczelaa@ornl.gov}
\affiliation{Department of Physics and Astronomy, University of Tennessee, Knoxville, Tennessee 37996, USA}
\affiliation{Neutron Scattering Division, Oak Ridge National Laboratory, Oak Ridge, Tennessee 37831, USA}

\author{H.~D.~Zhou}
\affiliation{National High Magnetic Field Laboratory, Florida State University, Tallahassee, Florida 32310-3706, USA}
\affiliation{Department of Physics and Astronomy, University of Tennessee, Knoxville, Tennessee 37996, USA}

\date{\today}

\begin{abstract}
Magnetoelectric materials are attractive for several applications, including actuators, switches, and magnetic field sensors. Typical mechanisms for achieving a strong magnetoelectric coupling are rooted in transition metal magnetism. In sharp contrast, here we identify CsEr(MoO$_{4}$)$_2$ as a magnetoelectric material without magnetic transition metal ions, thus ensuring that the Er ions play a key role in achieving this interesting property. Our detailed study includes measurements of the structural, magnetic, and electric properties of this material. Bulk characterization and neutron powder diffraction show no evidence for structural phase transitions down to 0.3~K and therefore CsEr(MoO$_{4}$)$_2$ maintains the room temperature {\it P2/c} space group over a wide temperature range without external magnetic field. These same measurements also identify collinear antiferromagnetic ordering of the Er$^{3+}$ moments below $T_{N}$ = 0.87 K. Complementary dielectric constant and pyroelectric current measurements reveal that a ferroelectric phase ($P \sim $ 0.5 nC/cm$^{2}$) emerges when applying a modest external magnetic field, which indicates that this material has a strong magnetoelectric coupling. We argue that the magnetoelectric coupling in this system arises from a pseudo Jahn-Teller distortion induced by the magnetic field. %Our work suggests that the search for new multiferroics should be expanded beyond conventional candidates.
\end{abstract}
\maketitle
\section{INTRODUCTION}
\begin{figure*}[tbp]
	\linespread{1}
	\par
	\begin{center}
		\includegraphics[width=0.9\textwidth]{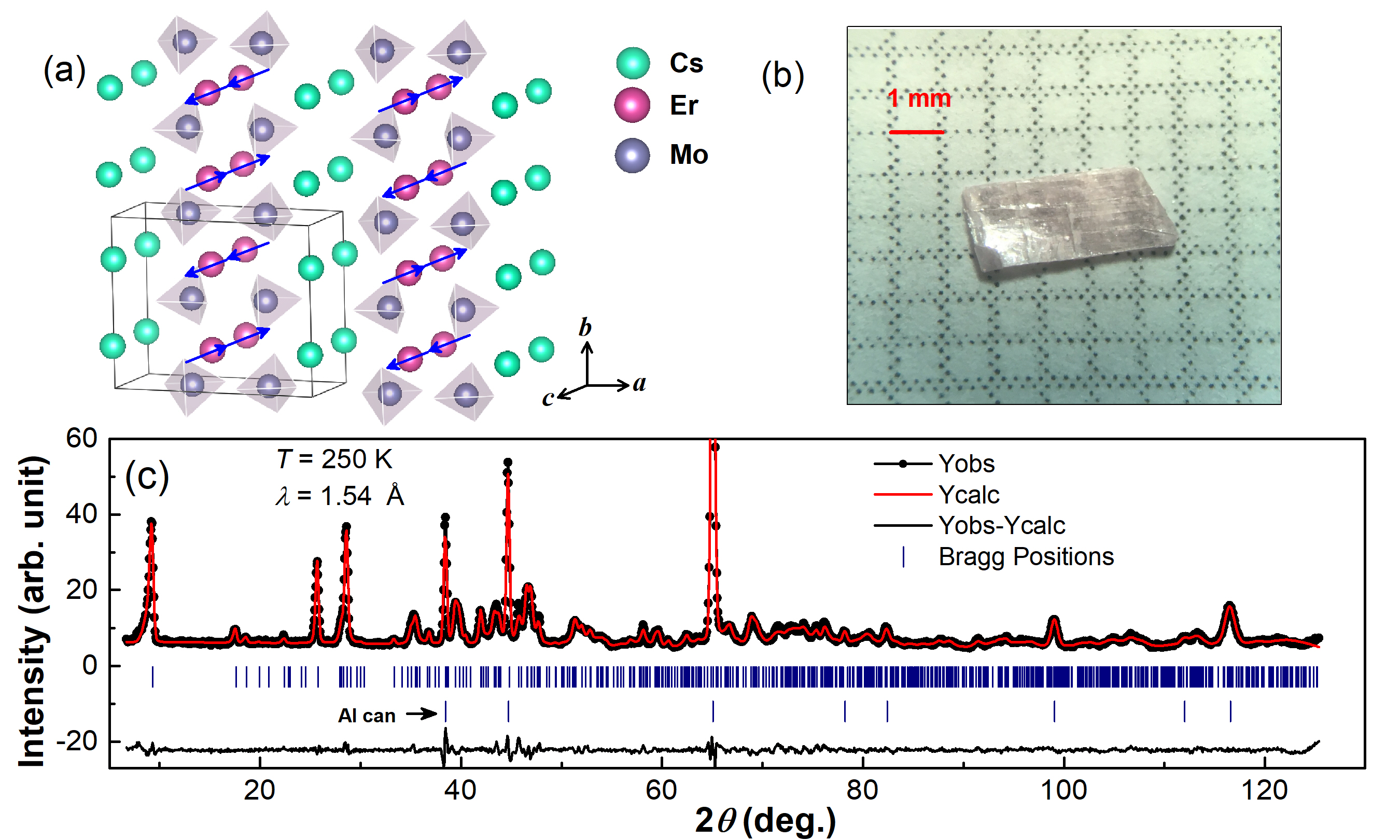}
	\end{center}
	\par
	\caption{\label{Fig1} (color online) (a) A schematic of the refined crystal and magnetic structures in the absence of an applied magnetic field. We do not show the O ions for clarity. (b) A picture of a representative CsEr(MoO$_4$)$_2$ single crystal used in this study. Several single crystals were crushed up for the HB-2A neutron powder diffraction experiment. (c) The refined neutron powder diffraction pattern collected at $T$ = 250 K. The data is plotted with black symbols, the structural model is superimposed on the data with a red curve, the ticks below the pattern show the expected Bragg peaks for the CsEr(MoO$_{4}$)$_{2}$ crystal structure and the Al sample can, and the difference pattern is indicated by a solid black curve.
}
\end{figure*}
The interest in magnetoelectric and multiferroic materials has been growing due to possible applications in new devices \cite{multireview1,multireview2}. Many efforts have focused on developing a detailed understanding of the magnetoelectric properties and mechanisms in materials based on magnetic transition metal ions. The strong magnetoelectric coupling in these systems has been achieved based on several design principles, including magnetostriction \cite{magnetostriction1,magnetostriction2,magnetostriction3}, spin current \cite{spincurrent1} and the inverse Dzyaloshinskii-Moriya interaction \cite{inverseDM1}. In most cases, a magnetic-field-induced modification of the ordered spin structure gives rise to changes in the ferroelectric properties \cite{hexa1}. Well-known multiferroics with a strong magnetoelectric coupling include the rare-earth manganites with general formula $R$MnO$_{3}$, where $R$ is a rare-earth element\cite{RMnO31}. However, it is generally argued that the role of the rare earth elements in this family is limited to a fine tuning of the exchange parameters arising from their slightly different ionic radii \cite{RMnO32}.

It was proposed recently that Jahn-Teller distortions (JTDs) can also provide a viable mechanism for achieving multiferroicity with an intrinsic magnetoelectric coupling\cite{multifromJTD1,multifromJTD2}. JTDs are generally active for non-linear molecules with degenerate electronic ground states \cite{JTD1}. JTDs occurring cooperatively via an electron-phonon coupling mechanism generate structural instabilities that are responsible for numerous intriguing and important physical phenomena such as high-T$_{c}$ superconductivity \cite{HTCSC1, HTCSC2, HTCSC3, HTCSC4} and colossal magnetoresistance \cite{CMR1,CMR2,CMR3,CMR4}. Although conventional JTDs are realized in systems with degenerate electronic energy levels, two non-degenerate levels separated by a small energy gap can also induce a JTD-like instability called a pseudo Jahn-Teller distortion (pJTD) when the energy gain from the deformation is larger than the gap. For a conventional JTD, group theoretical analysis shows that the normal mode associated with the distortion always has an even representation and therefore centrosymmetry is preserved if it is already present in the higher-symmetry state. For this reason, the conventional JTD can induce multiferroicity in non-centrosymmetric systems only \cite{multifromJTD1,multifromJTD2}, where the local electric dipole moments and magnetic ordering typically have different onset temperatures (i.e. Type I multiferroics) \cite{multifromJTD2}, which often results in a weak magnetoelectric coupling. On the other hand, this rule does not extend to the pJTD, as the two closely-spaced energy levels may have opposite parity and therefore can be mixed to make an odd phonon mode active as well. Consequently, a unique feature of a pJTD is its ability to induce local electric dipole moments in centrosymmetric systems \cite{pJTD1} and therefore this phenomenon is responsible for many common ferroelectrics \cite{pJTD2, pJTD3}. Interestingly, several recent theoretical studies have shown that a pJTD can also induce multiferroicity with a sizable magnetoelectric coupling \cite{pJTD3,pJTD_multi1,pJTD_multi2,pJTD_multi3}.

Although predicting and identifying JTD-induced multiferroic and magnetoelectric materials has been actively pursued, the focus again has mostly been limited to materials with magnetic moments arising from 3$d$ transition metals. Well-known examples include (Ga,Ge)(V,Mo)$_{4}$S$_{8}$ \cite{GVO1,GVO2,GVO3} and Ba$_{2}${\it T}Ge$_{2}$O$_{7}$ ({\it T} = V, Ni) \cite{multifromJTD2}. On the other hand, it has been known for a long time that rare-earth systems with local magnetic moments often exhibit structural phase transitions that originate from a JTD \cite{rareearthJTD1,rareearthJTD2,rareearthJTD3,rareearthJTD4}, but the possibility of multiferroicity and magnetoelectric coupling has been rarely studied. In contrast to 3$d$ transition metal systems, highly-localized electrons in rare-earth magnets generate weakly-dispersive bands, leading to large peaks in the electronic density-of-states. If these electronic bands can be tuned by external parameters such as a magnetic field so they exhibit significant overlap with peaks in the phonon density-of-states, and electron-phonon coupling is allowed by symmetry \cite{ARDM_review1}, then strong electron-phonon interactions may result.  

In this article, we report the discovery of a strong magnetoelectric coupling in a material where rare-earth elements are solely responsible for both magnetic order and ferroelectricity. We observed this magnetoelectric behavior below $T_{N}$ = 0.87 K in the alkali rare-earth double molybdate (ARDM) CsEr(MoO$_4$)$_2$, and we argue that it arises from a field-induced pJTD mechanism. Our comprehensive study shows that intrinsic magnetoelectric coupling may be ubiquitous in rare-earth compounds susceptible to JTDs and therefore calls for detailed experimental and theoretical studies to identify other magnetoelectric ARDMs. Specific investigations focused on elucidating the nature of the electronic energy levels and phonon modes involved in the pJTD for CsEr(MoO$_4$)$_2$ are also highly desirable.

\section{EXPERIMENTAL METHODS}
Single crystals of CsEr(MoO$_4$)$_2$ were grown by a flux method. Starting materials Cs$_2$CO$_3$, Er$_2$O$_3$ and MoO$_3$ with purities no less than 99.9\% were mixed with masses of 0.814\,g, 0.191\,g and 1.152\,g respectively in a covered alumina crucible. The mixture was then warmed up to 1000\textdegree C with a ramping rate of 100\textdegree C/hr and held there for 10 hours before cooling down to 600\textdegree C at a rate of 3\textdegree C/hour and then quenching to room temperature. The CsEr(MoO$_{4}$)$_{2}$ single crystals were isolated by washing the product with water.

High-resolution neutron powder diffraction (NPD) measurements were performed using the HB-2A powder diffractometer at the High Flux Isotope Reactor (HFIR) of Oak Ridge National Laboratory (ORNL). About 3 g of crushed single crystals were loaded in a cylindrical Al can and mounted in the He-3 insert of a cryostat to achieve a base temperature of 0.3~K. We used neutron wavelengths $\lambda$ = 1.54 and 2.41~\AA~with a collimation of open-21'-12'. We performed Rietveld refinements of the NPD patterns with the FullProf software suite \cite{Fprof}. 

The dc magnetic susceptibility data were obtained with a Quantum Design superconducting quantum interference device (SQUID) magnetometer. The high field dc magnetization was measured using a vibrating sample magnetometer (VSM) at the National High Magnetic Field Laboratory. The ac susceptibility was measured with the conventional mutual inductance technique with a homemade setup\cite{AC}. The specific heat data were obtained using a commercial Physical Property Measurement System (PPMS, Quantum Design) by the relaxation method. 

The capacitance was measured on single crystal samples with thin plate geometry by painting two faces with silver so they could be used as electrodes for an Andeen-Hagerling AH-2700A commercial capacitance bridge. The capacitance was converted to the dielectric constant by approximating the sample as an infinite parallel plate capacitor. To measure the electric polarization, we first measured the pyroelectric current under different cooling protocols and then integrated the pyroelectric current with respect to time. The detailed procedure of the pyroelectric current measurement has been published elsewhere \cite{Ipmeasurement}. 

\section{RESULTS}
We performed neutron powder diffraction (NPD) on crushed single crystals to determine the crystal structure at 250 K in the absence of an applied magnetic field. A schematic of the crystal structure is shown in \autoref{Fig1}(a) and a picture of a representative single crystal is depicted in \autoref{Fig1}(b). The crystal has a thin-plate geometry, is easily cleaved, and the normal to the flat face is parallel to the $a$-axis. The Rietveld refinement of the 250~K data is presented in \autoref{Fig1}(c) and reveals that the Bragg peaks can be indexed with the monoclinic space group {\it P2/c}, which is consistent with the previous room-temperature structure determination \cite{94_khatsko}. The low crystal symmetry likely arises from severe distortions of the orthorhombic KEr(MoO$_4$)$_2$ structure due to the large ionic radius of Cs \cite{ARDM_review1}. Notably, structural chains of Er$^{3+}$ ions run along the {\it b}-axis, which may be indicative of a low-dimensional magnetic system. This arrangement of magnetic ions is in sharp contrast to the trigonal system RbFe(MoO$_4$)$_2$, where the magnetic Fe ions form a frustrated triangular sublattice that generates a 120 degree spin structure \cite{07_kenzelmann}.

We collected additional NPD data with $\lambda$ = 1.54~\AA~down to 1.5~K to look for changes in the diffraction pattern that would be indicative of structural phase transitions, but none were found. Therefore, CsEr(MoO$_{4}$)$_2$ retains the {\it P2/c} space group over the entire temperature range investigated. The lattice constants, fractional coordinates, and key structural parameters at 1.5~K and 250~K are presented in \autoref{Table1}. Notably, the Wyckoff position of the magnetic ion Er$^{3+}$ is 2f and its point symmetry is $C_{2}$, ensuring a common quantization axis $z$ for all Er$^{3+}$ ions along the {\it c}-axis. According to Hund's rules, the Er$^{3+}$ ground-state multiplet is described by $J = \frac{15}{2}$. Due to the low local symmetry of Er$^{3+}$, crystalline electric fields (CEFs) from the neighboring oxygen ions are expected to split the $2J+1$ degenerate electronic levels into eight Kramers doublets. Previous work has shown that the ground state doublet and first excited state have predominantly $\left|J_z = \pm \frac{15}{2}\right>$ and $\left|J_z = \pm \frac{13}{2}\right>$ character, respectively \cite{94_khatsko}. In other words, CsEr(MoO$_{4}$)$_2$ is an Ising magnet with a strong easy axis anisotropy along the crystalline {\it c}-axis \cite{note}. 

\begin{table}[htb]
\begin{center}
\caption{Structural parameters for CsEr(MoO$_4$)$_2$ extracted from the refinements of the~1.54~\AA~neutron powder diffraction data. The lattice constants and Er-Er intrachain distance are in \AA~and the $\beta$ angle is in degrees.} 
\label{Table1}
\begin{tabular}{l l l l l}
\hline 
\hline
$T$ & 250~K & 1.5~K \\
\hline
SG & {\it P2/c} & {\it P2/c} \\
$a$ & 9.5102(9) & 9.470(1) \\  
$b$ & 7.9196(7) & 7.9062(8) \\
$c$ & 5.0350(3) & 5.0295(4) \\
$\beta$ & 91.265(7) & 91.374(8) \\
Er & 0.5, 0.25, 0.006(3) & 0.5, 0.25, 0.003(3)  \\ 
Cs & 0, 0.25, 0.040(3) & 0.5, 0.25, 0.035(3) \\ 
Mo & 0.309(1), -0.004(1), 0.519(1) & 0.310(1), -0.003(2), 0.522(1)  \\
O$_1$ & 0.437(1), 0.000(1), 0.251(1) & 0.439(1), 0.004(1), 0.250(1) \\
O$_2$ & 0.307(1), 0.175(1), 0.746(2) & 0.305(1), 0.176(1), 0.746(2) \\
O$_3$ & 0.140(1), 0.982(1), 0.387(1) & 0.138(1), 0.982(1), 0.385(1) \\
O$_4$ & 0.313(1), 0.815(1), 0.732(2) & 0.311(1), 0.813(1), 0.733((2) \\
Er-Er & 3.96 & 3.95 \\
R$_\mathrm{wp}$ & 5.61~\% & 6.79~\%  \\ 
$\chi^2$ & 7.60 & 11.2 \\
\hline\hline
\end{tabular}
\end{center}
\end{table}

%%%%%$Explanation for structure$
\begin{figure}[tbp]
	\linespread{1}
	\par
	\begin{center}
		\includegraphics[width=0.5\textwidth]{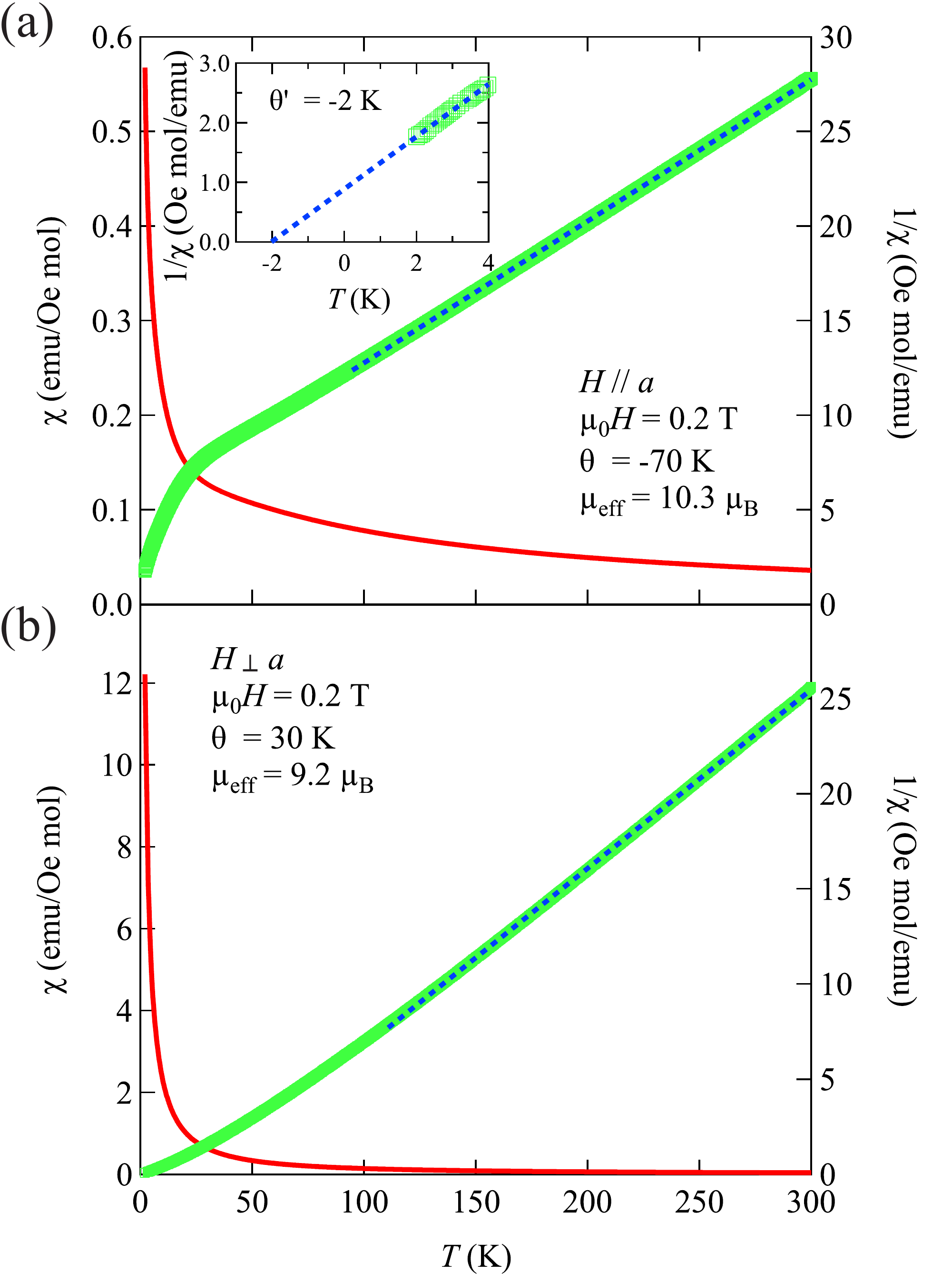}
	\end{center}
	\par
	\caption{\label{Fig2} (color online) The temperature-dependence of the dc susceptibility and its inverse for CsEr(MoO$_4$)$_2$ with (a) $H$ // $a$-axis and (b) $H$ $\perp$ $a$ axis. The dashed lines represent fits to the Curie-Weiss law. The inset shows the inverse susceptibility vs temperature over the low-temperature range only with the corresponding Curie-Weiss law fit superimposed on the data.
}
\end{figure}

As shown in \autoref{Fig2}, the temperature-dependence of the dc susceptibility for CsEr(MoO$_4$)$_2$, $\chi (T)$, exhibits no signs of magnetic ordering down to 2 K for both $H$ // $a$-axis and $H$ $\perp$ $a$-axis. A linear Curie-Weiss fit of the inverse dc susceptibility data from 100-300 K yields $\mu_{\rm eff}\!=\!10.3$~$\mu_{\rm B}$ and $\theta_{\rm CW}\!=\!-70$~K for $H$ // $a$-axis and $\mu_{\rm eff}\!=\!9.2$~$\mu_{\rm B}$ and $\theta_{\rm CW}\!=\!30$~K for $H$ $\perp$ $a$-axis. These effective moment values are consistent with the free ion value of $\mu_{\rm eff}\!=\!9.58$~$\mu_{\rm B}$ for Er$^{3+}$ ($^4$I$_{15/2}$). The dc magnetic susceptibility for $H$ $\perp$ $a$-axis is about 20 times higher than for $H$ // $a$-axis in the paramagnetic regime, indicative of a hard axis along the $a$-direction. The large $\theta_{\rm CW}$ temperatures are not necessarily indicative of strong exchange between Er$^{3+}$ ions because the intrinsic values may be significantly influenced by low-lying crystal field levels. Furthermore, the sign difference of $\theta_{\rm CW}$ for the two directions likely arises from the high degree of magnetic anisotropy in this system, which is a typical characteristic of rare-earth magnets. The inset in \autoref{Fig2}(a) shows a Curie-Weiss fit of the inverse susceptibility over the low temperature range only, and the resulting $\theta_{\rm CW}' =$~-2 K is likely more indicative of the true exchange interaction strength between Er$^{3+}$ ions. Close agreement between $\left|\theta_{\rm CW}'\right|$ and the ordering temperature identified in the low-temperature dc susceptibility and specific heat measurements described below indicate no strong magnetic frustration in this system \cite{frustration1}, which is consistent with the arrangement of the Er$^{3+}$ ions in the crystal structure.
\begin{figure}[tbp]
	\linespread{1}
	\par
	\begin{center}
		\includegraphics[width=0.5\textwidth]{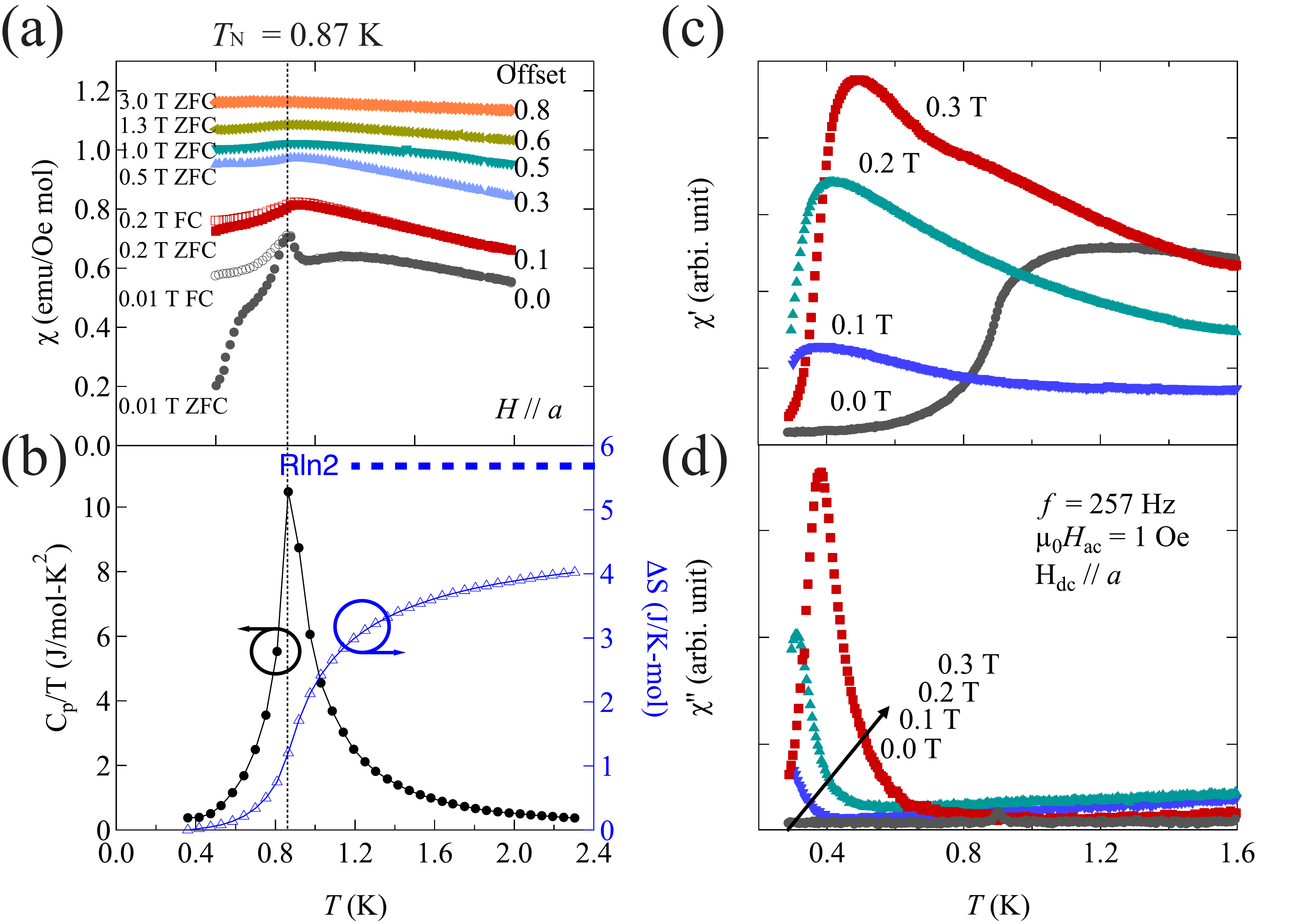}
	\end{center}
	\par
	\caption{\label{Fig3} (color online) (a) The temperature-dependence of the dc susceptibility below 2 K with $H$ // $a$-axis for different applied magnetic fields. The data collected at different fields is vertically-offset for clarity. (b) The specific heat $C_p$, plotted as $C_p/T$, and the entropy recovered on warming as a function of temperature. (c) The real part of the ac susceptibility vs temperature for different dc fields $H$ // $a$-axis. (d) The imaginary part of the ac susceptibility vs temperature for different dc fields $H$ // $a$-axis. }
\end{figure}

The dc susceptibility measured below 2 K with $H$ // $a$-axis is shown in \autoref{Fig3}(a); for $\mu_0H$ = 0.01 T a broad peak is observed around 1.2 K followed by a sharp peak and the onset of a zero-field-cooled (ZFC) / field-cooled (FC) divergence at 0.87 K. Although the broad peak disappears and the sharp peak becomes less pronounced with increasing field, the latter is still clearly visible and essentially temperature-independent for $\mu_0H$~$<$~3~T. A sharp lambda anomaly is also observed in the specific heat data shown in \autoref{Fig3}(b) at 0.87~K, which is consistent with the dc susceptibility data and therefore likely indicative of long-range magnetic order. Notably, the entropy recovered when warming through the phase transition is less than $R$ln2 expected for a rare earth magnet with a CEF ground state doublet. The real part of the ac susceptibility measured with zero dc field, shown in \autoref{Fig3}(c), displays a broad peak around 1.2 K and a sharp drop around 0.87 K; these observations are in good agreement with the dc susceptibility and specific heat data. The imaginary part of the ac susceptibility measured with zero dc field, presented in \autoref{Fig3}(d), exhibits a small sharp peak around 0.87 K and therefore provides further support for long-range magnetic ordering below this temperature. The long-range ordering transition temperature identified here agrees well with previous work \cite{94_khatsko}.
\begin{table}[h]
	\begin{tabular}{ccc|cccccc}
		IR  &  BV  &  Atom & \multicolumn{6}{c}{BV components}\\
		     &        &           & $m_{\|a}$ & $m_{\|b}$ & $m_{\|c}$ &$im_{\|a}$ & $im_{\|b}$ & $im_{\|c}$ \\
		\hline
		$\Gamma_{1}$ & $\psi_{1}$  &      1 &      1 &      0 &      0 &      0 &      0 &      0  \\
		                        &                  &      2 &     -1 &      0 &      0 &      0 &      0 &      0  \\
		                        & $\psi_{2}$ &      1 &      0 &      1 &      0 &      0 &      0 &      0  \\
		                        &                  &      2 &      0 &     -1 &      0 &      0 &      0 &      0  \\
		$\Gamma_{2}$ & $\psi_{3}$  &      1 &      1 &      0 &      0 &      0 &      0 &      0  \\
		                        &                  &      2 &      1 &      0 &      0 &      0 &      0 &      0  \\
		                        & $\psi_{4}$  &      1 &      0 &      1 &      0 &      0 &      0 &      0  \\
		                        &                   &      2 &      0 &      1 &      0 &      0 &      0 &      0  \\
		$\Gamma_{3}$  & $\psi_{5}$  &      1 &      0 &      0 &      1 &      0 &      0 &      0  \\
		                         &                  &      2 &      0 &      0 &     -1 &      0 &      0 &      0  \\
		$\Gamma_{4}$  & $\psi_{6}$  &      1 &      0 &      0 &      1 &      0 &      0 &      0  \\
		                        &                   &      2 &      0 &      0 &       1 &     0 &      0 &      0  \\
	\end{tabular}
	\caption{Basis vectors (BV) for possible irreducible representations of the space group {\it P2/c} with ${\bf k} = (0.5, 0, 0)$ and Er site (0.5, 0.25, 0.003).}
	\label{Table2}
\end{table}

The missing entropy when warming above $T_N$ and the broad peaks in the dc magnetic susceptibility and real part of the ac magnetic susceptibility are hallmarks of a low-dimensional magnetic system. Interestingly, a simple examination of the crystal structure reveals Er chains running along the {\it b}-axis and therefore supports quasi-one-dimensional behavior, although a power law fit of the magnetic Bragg peak intensity for the related compound CsDy(MoO$_4$)$_2$ is indicative of quasi-two-dimensional Ising behavior instead \cite{04_khatsko}. Regardless of the specific nature of the low-dimensionality, the broad peaks observed in the susceptibility measurements indicate that short-range correlations are building up with decreasing temperature before the onset of long-range magnetic order. The low-dimensionality of this system may also explain the ZFC/FC divergence in the dc magnetic susceptibility below $T_N$, as finite correlation lengths along one or both interchain directions can produce magnetic domains that can be manipulated by temperature or an applied magnetic field. Previous support for this idea comes from calculating possible magnetic structures stabilized by dipolar interactions, as it is found that there are two antiferromagnetic chain configurations with different interchain correlations that are comparable in energy \cite{94_khatsko}. Previous work on the related compound CsDy(MoO$_4$)$_2$ is also relevant, as a finite correlation length along the $a$-direction was identified below $T_N$ that could be tuned by the cooling rate through $T_N$ \cite{04_khatsko}. This magnetic domain formation and rearrangement under magnetic field may be responsible for the peaks observed in the the ac susceptibility data when a small dc field is applied.
\begin{figure*}[tbp]
	\linespread{1}
	\par
	\begin{center}
		\includegraphics[width=1\textwidth]{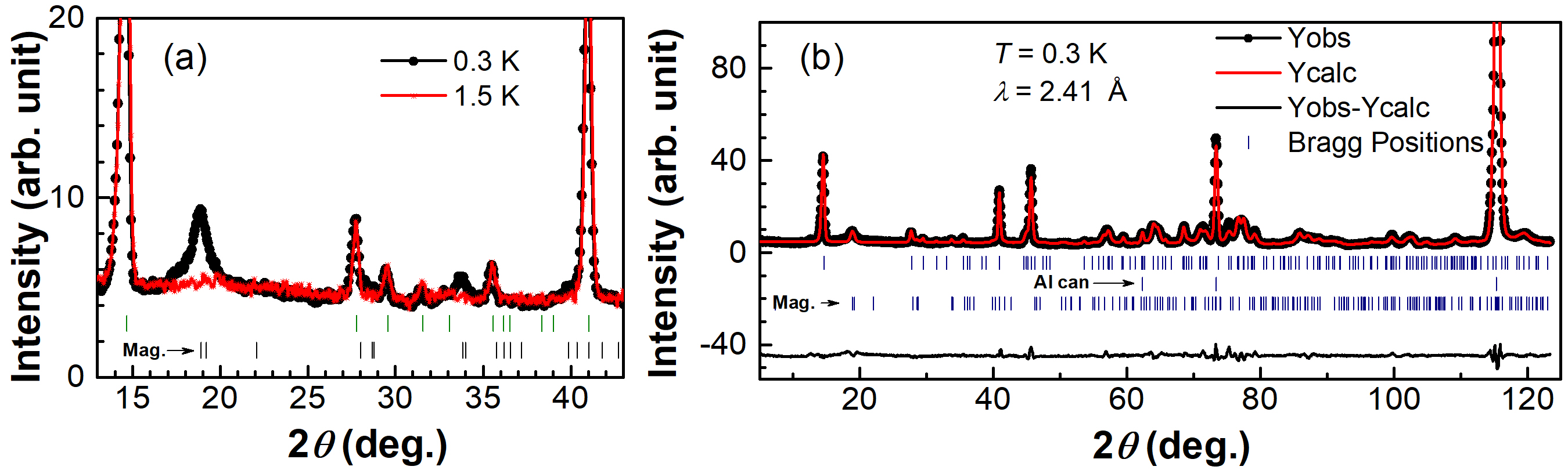}
	\end{center}
	\par
	\caption{\label{Fig4} (color online) (a) Neutron powder diffraction patterns collected at 1.5 K (red curve) and 0.3 K (black symbols). (b) The refined neutron powder diffraction pattern collected at $T$ = 0.3 K. The data is plotted with black symbols, the refinement result is superimposed on the data with a red curve, the ticks below the pattern show the expected Bragg peaks for the CsEr(MoO$_{4}$)$_{2}$ crystal structure, the Al sample can, and the CsEr(MoO$_{4}$)$_{2}$ magnetic structure, and the difference pattern is indicated by a solid black curve. 
}
\end{figure*}

\begin{figure}[tbp]
	\linespread{1}
	\par
	\begin{center}
		\includegraphics[width=0.5\textwidth]{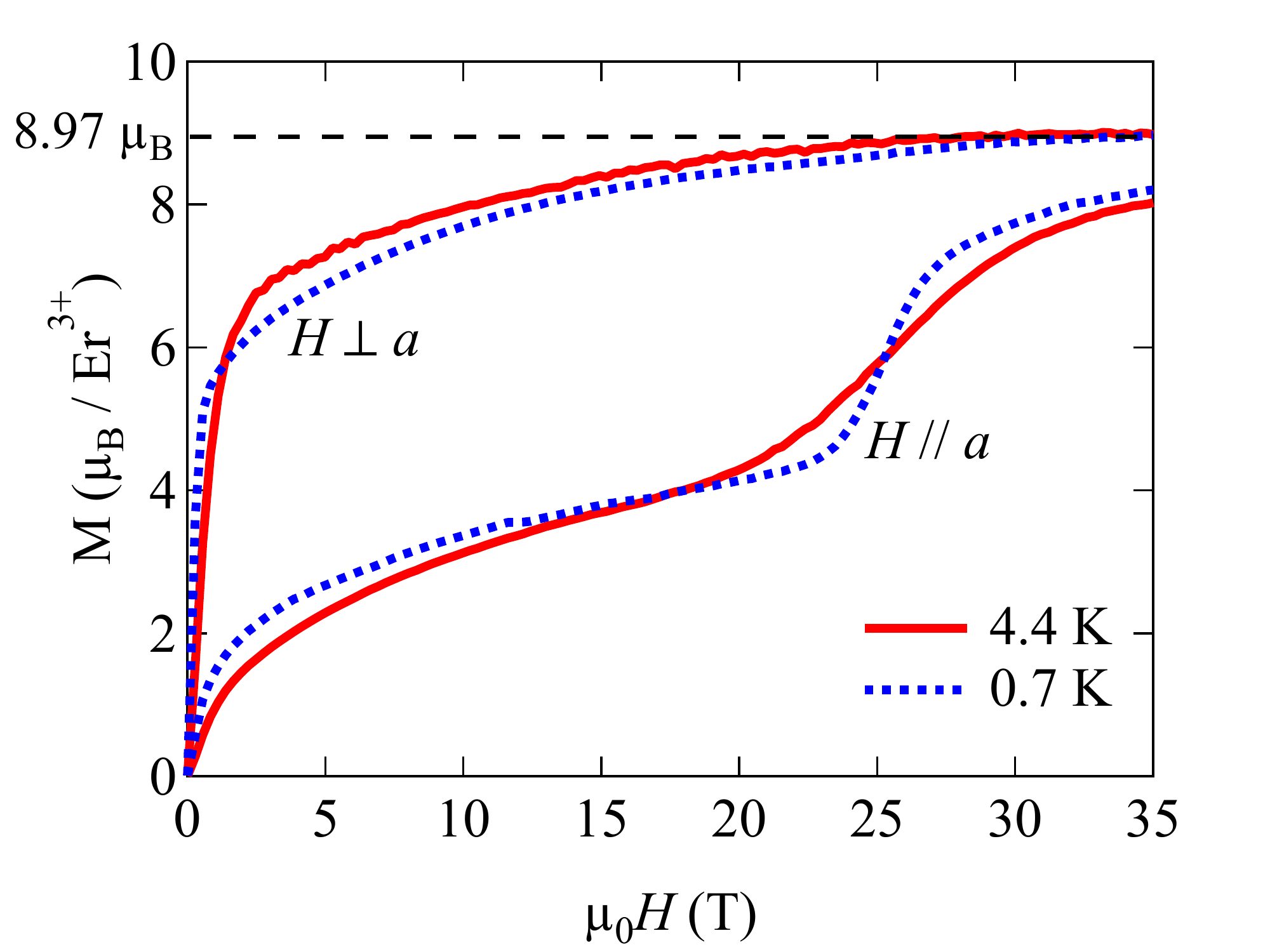}
	\end{center}
	\par
	\caption{\label{Fig5} (color online) The magnetization measured as a function of applied magnetic field at 0.7 K and 4.4 K for $H$ // $a$-axis and H $\perp$ $a$-axis.
}
\end{figure}

To determine the magnetic structure below $T_{N}$ in the absence of an applied magnetic field, we collected neutron powder diffraction data at 0.3 and 1.5 K using the HB-2A powder diffractometer with a neutron wavelength of 2.41~\AA. We verified that CsEr(MoO$_4$)$_2$ remained in the $P2/c$ space group at 0.3~K and we identified magnetic Bragg peaks in the same dataset below $T_N$. Interestingly, the magnetic Bragg peaks are not resolution-limited, which is most apparent for the extra scattering centered at $2\theta$~$=$~19\textdegree~as shown in \autoref{Fig4}(a). The broad nature of these peaks is likely a consequence of a finite correlation length along one or more interchain directions, as discussed above. Nonetheless, the magnetic peaks observed can be indexed with the propagation vector ${\bf k}$~$=$~(0.5 0 0). To model the magnetic structure we first performed a symmetry analysis using SARAh \cite{00_wills}. Ultimately, we find that the best magnetic refinement of the 0.3\,K diffraction pattern is achieved by using the $\Gamma_{3}$ irreducible representation (see~\autoref{Table2} for details). The refined 0.3~K diffraction pattern, shown in \autoref{Fig4}(b), yields an ordered Er moment of $m_c=2.80(3)\,\mu_B$. Although the ordered moment magnitude is much smaller than the value of $\sim$~9~$\mu_B$ expected for a $\left|J_z = \pm \frac{15}{2}\right>$ CEF ground state doublet, this is not surprising because the refinement assumes that the magnetic Bragg peaks are resolution-limited. The spin structure is schematically presented in \autoref{Fig1}(a) and consists of antiferromagnetic chains of Er ions running along the {\it b}-axis with the moments pointing along the {\it c}-axis. We note that this moment direction is consistent with the Ising anisotropy determined for this system previously \cite{94_khatsko, note}. The magnetic structure of CsEr(MoO$_4$)$_2$ also consists of interchain correlations that are ferromagnetic along the {\it c}-axis and antiferromagnetic along the $a$-axis, and it is reminiscent of the ordered spin configuration for the related system KTb(WO$_4$)$_2$ \cite{06_loginov}. This spin arrangement corresponds to the expected magnetic ground state for a simple Hamiltonian with nearest-neighbor dipolar interactions only \cite{94_khatsko}.

We now turn to the magnetic properties of CsEr(MoO$_4$)$_2$ in the presence of significant applied magnetic fields. \autoref{Fig5} shows the isothermal magnetization measured with both $H$ // $a$-axis and $H$ $\perp$ $a$-axis at 0.7 K and 4.4 K. The strong directional-dependence of this data provides additional support for the large Ising anisotropy of this system. For $H$ $\perp$ $a$-axis, the magnetization sharply increases below 0.5 T and gradually approaches a saturation value of 8.97 $\mu_{B}$, which is the moment size for a free Er$^{3+}$ ion. Below $T_{N}$, similar features are observed with a slightly sharper increase at low fields. The relatively low applied fields required for $H$ $\perp$ $a$-axis to saturate the magnetization to the Er$^{3+}$ free ion value are consistent with a CEF ground state doublet of predominantly $\left|J_z = \pm \frac{15}{2}\right>$ character both above and below $T_N$, as identified in previous work \cite{94_khatsko}. Although we find no clear evidence of metamagnetic transitions up to 35~T for this field orientation, previous work indicates that the sharp increase in low fields arises from a metamagnetic transition below $T_N$ for $H$ // {\it c}-axis \cite{94_khatsko}. Presumably, this corresponds to a spin flip transition within the Er chains. Our high-field data cannot resolve this transition clearly due to the coarse point density in the low-field regime and the lack of in-plane crystal orientation information. On the other hand, our magnetization curves at 4.4 K and 0.7 K for $H$ // $a$-axis show a clear upturn centered at 25~T, which is indicative of a distinct metamagnetic transition. Since this transition persists well above $T_N$, we anticipate that it arises from crystalline electric field effects. More specifically, there is likely a crossing of crystal field levels at 25~T due to a larger {\it g}-factor value along the $a$-axis for the excited state relative to the ground state doublet. We also note that since the isothermal magnetization curves above and below $T_N$ are similar along both directions, the internal field from the magnetic ordering only generates a small splitting of the CEF Kramers doublets below $T_N$, which is typical for rare-earth magnets \cite{REintermetallic1}.
\begin{figure}[tbp]
	\linespread{1}
	\par
	\begin{center}
	    \includegraphics[width=0.5\textwidth]{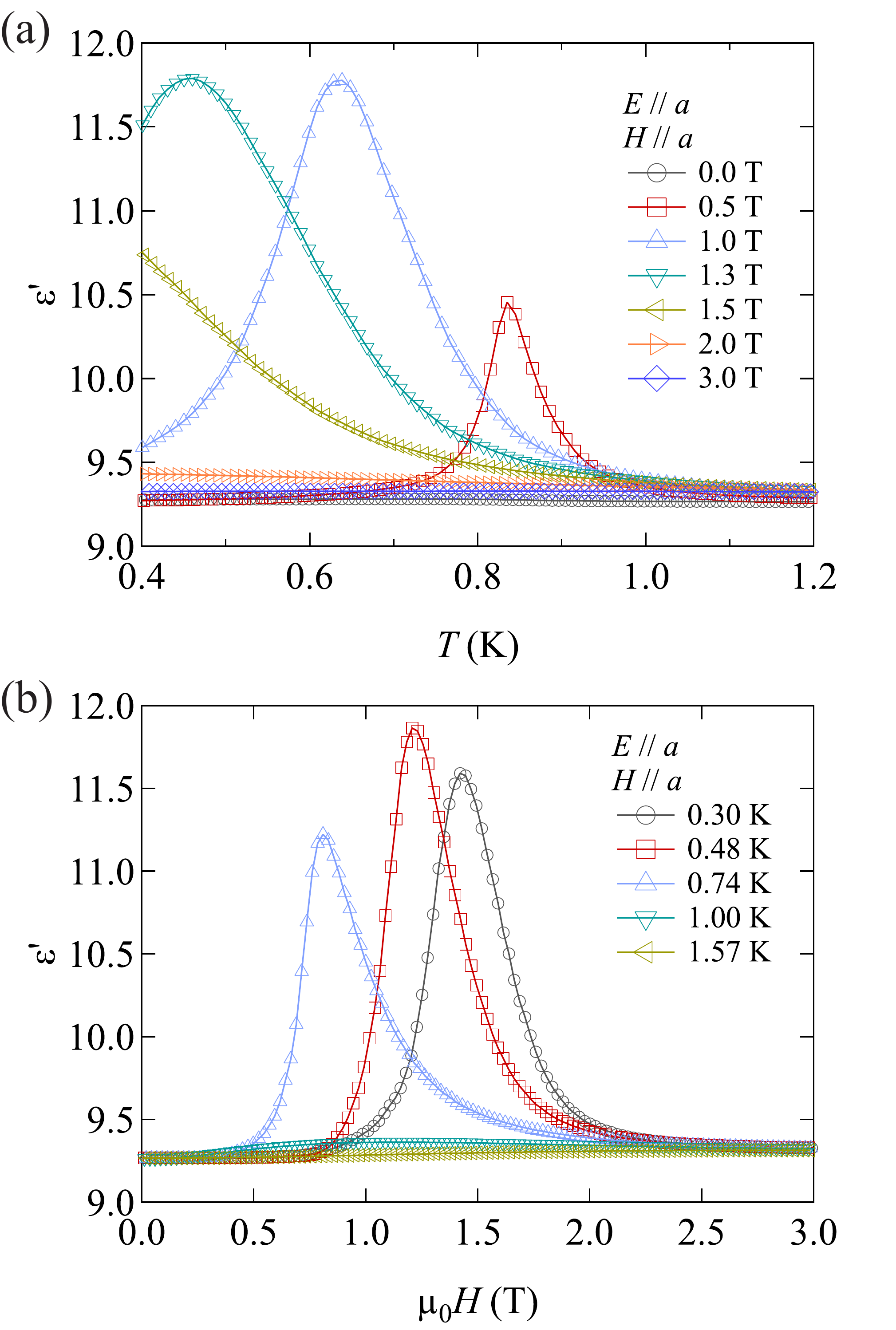}
	\end{center}
	\par
	\caption{\label{Fig6} (color online) (a) Temperature and (b) magnetic field dependence of the dielectric constant with $E$ // $a$-axis and $H$ // $a$-axis. An anomaly is observed for applied magnetic fields between 0.5 and 2 T at temperatures below $T_N$. 
}
\end{figure}

\autoref{Fig6}(a) shows the temperature and magnetic field dependence of the dielectric constant for CsEr(MoO$_4$)$_2$ with {\it E} // $a$-axis, which will now be referred to as $\varepsilon' (T)$ and $\varepsilon' (H)$ respectively. $\varepsilon' (T)$ shows no obvious anomaly below 1.5 K in the absence of an applied magnetic field. On the other hand, with $\mu_0 H$ = 0.5 T applied along the $a$-axis a peak is visible at 0.84 K, which is slightly lower than $T_{N}$. As the magnetic field increases to 1~T, the intensity of this peak increases and its position shifts to 0.64 K. As the field is increased further, this peak continuously shifts to lower temperatures and disappears by 3~T. Meanwhile, $\varepsilon' (H)$ measured at 0.3 K with $H$ // $a$-axis displays a sharp peak at $\mu_0 H$ = 1.5 T. As the temperature increases, the peak intensity reaches a maximum at 0.48 K before rapidly decreasing as the temperature rises above $T_{N}$. The dielectric constant is also highly anisotropic, as we found that $\varepsilon' (T)$ with $H$ $\perp$ $a$-axis is anomaly-free (not shown here). 
\begin{figure}[tbp]
	\linespread{1}
	\par
	\begin{center}
		\includegraphics[width=0.5\textwidth]{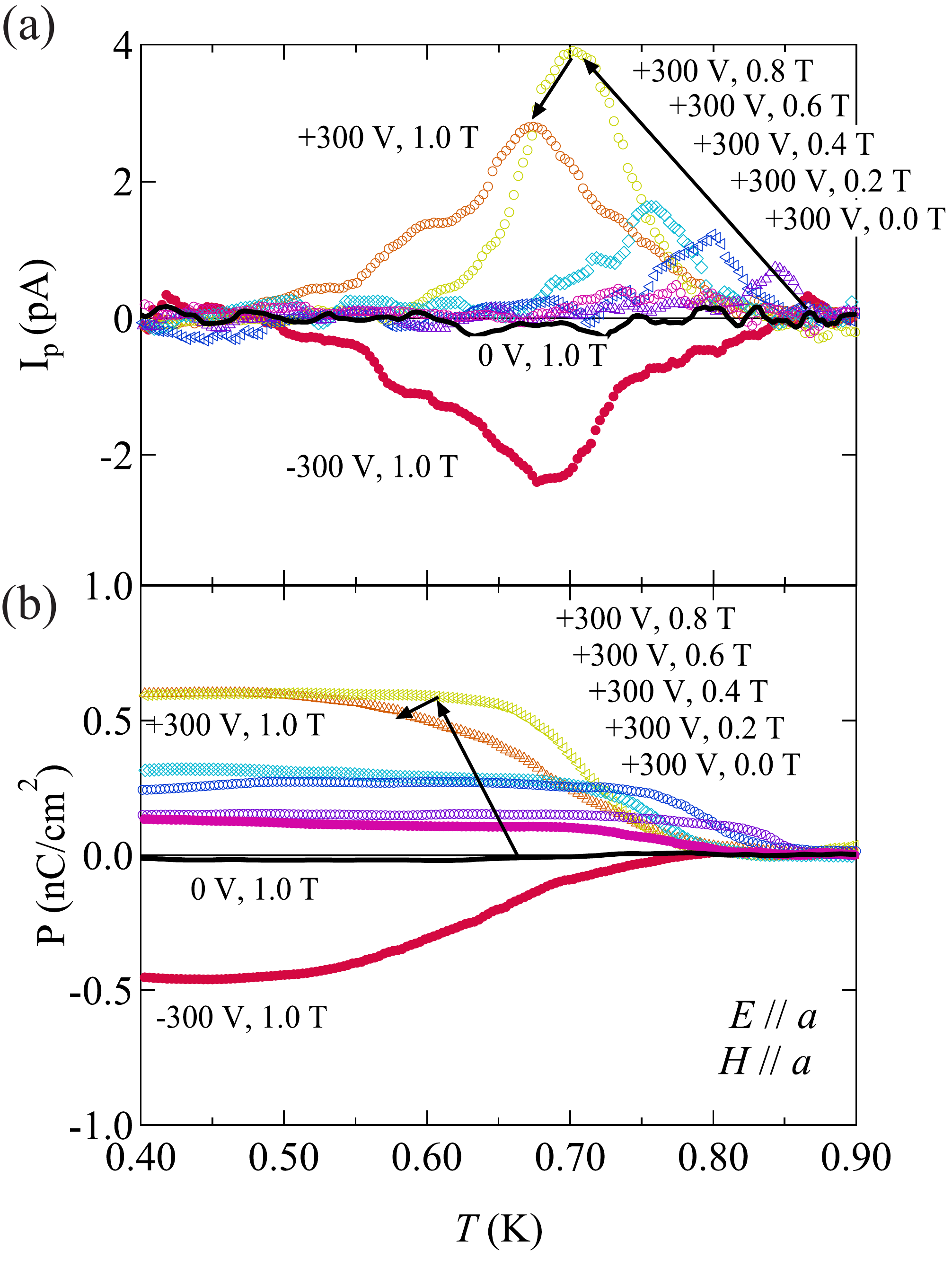}
	\end{center}
	\par
	\caption{\label{Fig7} (color online) (a) The pyroelectric current measured with $H$ // $a$-axis and $E$ // $a$-axis. (b) The electric polarization for $H$ // $a$-axis and $E$ // $a$-axis, obtained by integrating the pyroelectric current with time.}
\end{figure}

\autoref{Fig7}(a) shows the pyroelectric current measured at various magnetic fields ($H$) applied along the $a$-axis with various poling electric fields. The spontaneous and reversible pyroelectric current confirms the ferroelectricity of this compound below $T_{N}$. Although there is no obvious anomaly observed below 1 K in the absence of a magnetic field, a modest applied field of 0.2 T generates a peak below 0.9 K. With increasing magnetic field up to 0.8~T, the peak intensity increases and its position shifts to lower temperatures. When $\mu_{0}H$ increases to 1 T, the pyroelectric current peak decreases slightly and is centered at 0.67~K, which is consistent with the 0.64~K peak in the dielectric constant observed at the same magnetic field. To check the switchability of the polarization, we also measured pyroelectric current under opposite poling electric field of -300 V with the magnetic field held fixed at 1 T, and indeed the sign of the pryoelectric current changes accordingly. Additionally, we measured the pyroelectric current without poling electric field applied and found no measurable response due to randomly-oriented ferroelectric domains. These features strongly suggest that ferroelectricity is induced with an applied magnetic field $H$ // $a$-axis. We also found
that the presence or absence of magnetic field during cooling
did not make any difference. \autoref{Fig7}(b) shows the electric polarization obtained by integrating the pyroelectric current with time. Although the polarization is tiny, the finite value demonstrates that the magnetic field-induced structural phase transition is associated with spontaneous electric dipole moments. Taken together with the evidence from dc magnetic susceptibility for long-range magnetic order with $H$ // $a$-axis up to 3 T (at least), this indicates that CsEr(MoO$_4$)$_2$ is a strong magnetoelectric material for particular combinations of temperature and magnetic field. 

\section{DISCUSSION}

\begin{figure}[tbp]
	\linespread{1}
	\par
	\begin{center}
		\includegraphics[width=0.5\textwidth]{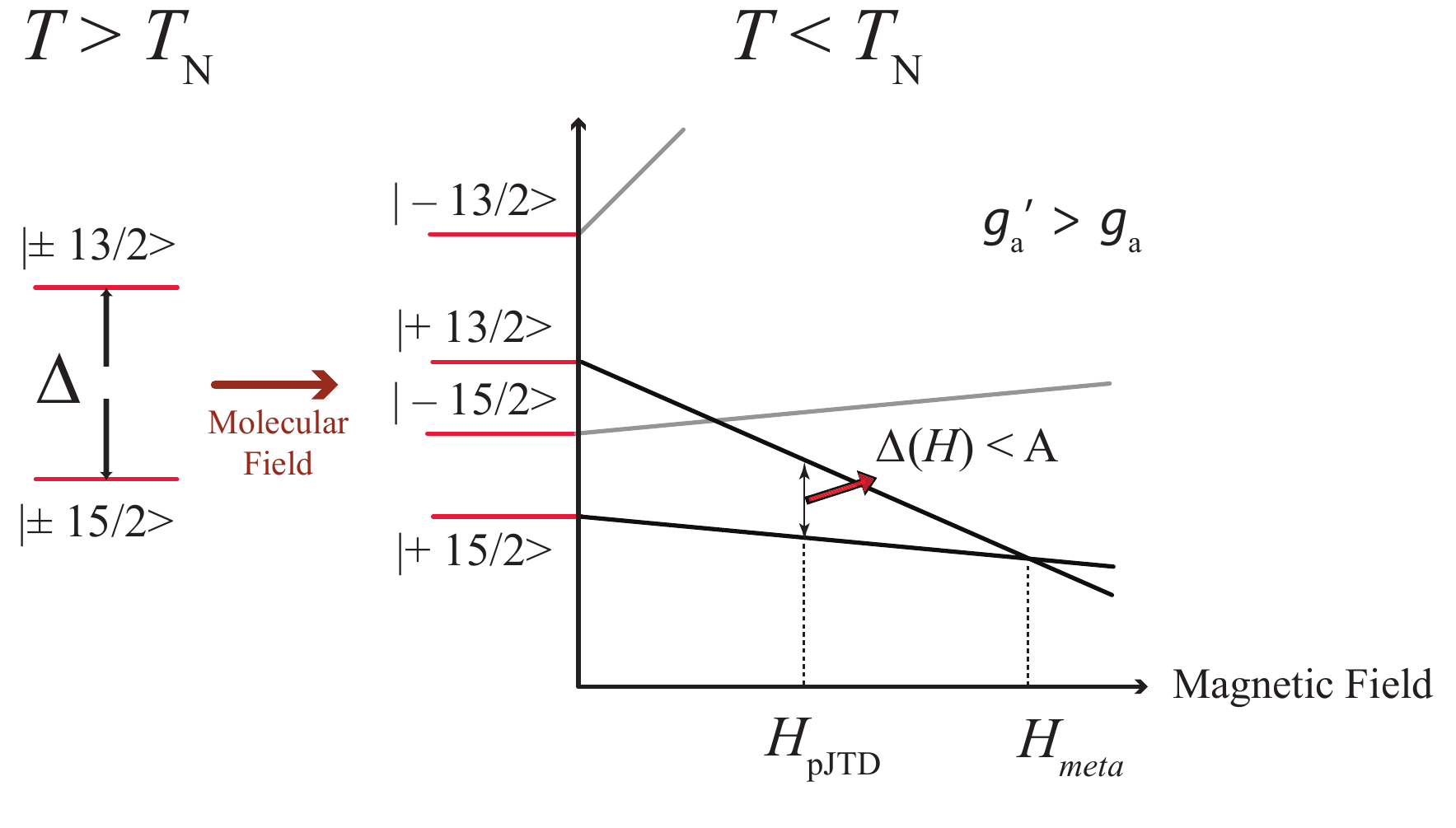}
	\end{center}
	\par
	\caption{\label{Fig8} (color online) Possible electronic energy level diagram for CsEr(MoO$_{4}$)$_{2}$. The CEF ground state doublet and first excited state have predominantly $\left|J_z = \pm \frac{15}{2}\right>$ and $\left|J_z = \pm \frac{13}{2}\right>$ character, respectively, and they are separated by an energy gap $\Delta$ of 1.6 meV \cite{96_khatsko}. The internal magnetic field generated by the magnetic order below $T_N$ only leads to a small splitting of the Kramers' doublets, but an applied magnetic field with $H$ // $a$-axis produces a ferroelectric transition at $H_{\text{pJTD}}$, where the condition $A>\Delta(H)$ is met. Increasing the magnetic field further generates a CEF level crossing and hence a metamagnetic transition at $H_{\text{meta}}$. The small value of $\Delta$ and the identification of the 25 T metamagnetic transition imply that a pseudo Jahn-Teller distortion is possible in this system.}
\end{figure}

Given the results from the various experimental techniques, we describe a possible electronic energy diagram in \autoref{Fig8} to explain how an applied magnetic field induces ferroelectricity in this compound. As discussed above, previous work for CsEr(MoO$_{4}$)$_{2}$ has identified a CEF ground state doublet and a first excited state that are close to pure $\left|J_z = \pm \frac{15}{2}\right>$ and $\left|J_z = \pm \frac{13}{2}\right>$, respectively \cite{94_khatsko}. Below $T_N$, the internal magnetic field generated by the ordered state leads to a small splitting of the Kramers' doublets but no CEF level crossings. In the presence of an increasing applied magnetic field with $H$ // $a$-axis, the larger g-factor for the first excited state (i.e. $g_{a}'$) relative to the ground state value $g_a$ make the difference in energy smaller. When the necessary condition $A > \Delta (H)$ is met for the pJTD, where $A$ is the distortion-induced energy gain and $\Delta(H)$ is the energy gap between the two states under field, the structural phase transition to the ferroelectric phase occurs. Increasing the magnetic field further generates the CEF level crossing at 25~T observed in the magnetization measurements. This field-induced sequence of transitions is consistent with the extremely low energy gap $\Delta$ = 1.6 meV between the CEF ground state doublet and first excited state \cite{96_khatsko}. Therefore, a pseudo Jahn-Teller distortion incorporating the  $\left|J_z = \frac{15}{2}\right>$ and $\left|J_z = \frac{13}{2}\right>$ CEF levels, as illustrated in \autoref{Fig8}, is a viable mechanism for inducing ferroelectricity in this material. %We speculate that CsEr(MoO$_{4}$)$_{2}$ is a Type I multiferroic as we currently have no evidence for an $H$ // $a$-axis metamagnetic transition that is coincident with the field-induced ferroelectric transition, but this topic requires further study. 

%We infer that this pseudo degeneracy is how the pseudo Jahn-Teller type structural instability can take place to the ferroelectric ground state although the magnetically ordered CsEr(MoO$_{4}$)$_{2}$ is not Jahn-Teller crystal due to the lack of degeneracy. However, controlling the external magnetic field can induce the pseudo degeneracy that can mix the phonon mode with different parity. Although there is a sharp increase in magnetization at low field below 1 T for all directions as \autoref{Fig5} shows, the ferroelectricity and dielectric anomalies do not drastically changes in contrast to other Type - 2 multiferroics. We note, therefore, that although magnetic long-range ordering and ferroelectricity coexists in this compound, the ferroelectricity is not strongly coupled to detail of the magnetic structure. This also supports for the Jahn-Teller distortion origin ferroelectricity in this compounds.
%COMMENT: I don't understand the point of this paragraph, other than speculating that our material is a Type I multiferroic. I added on sentence to the previous paragraph to get this point across. 

In highly anisotropic crystals such as CsEr(MoO$_{4}$)$_{2}$, the acoustic phonon dispersions can generate a peak in the phonon density-of-states that is similar in energy to the magnetic field-induced pseudo-degenerate CEF states \cite{ARDM_review1}. If this condition is met and spin-phonon coupling is allowed by symmetry, then strong spin-lattice coupling may arise. Recall that the electric polarization for CsEr(MoO$_{4}$)$_{2}$ appears parallel to the $a$-axis only. It is then interesting to note that previous work on the sister compound CsDy(MoO$_{4}$)$_{2}$ identified acoustic phonons with wavevectors $\vec{k}$~$\parallel$~$a$-axis as the driving force for the co-operative Jahn Teller structural transition at 38 K \cite{phononmode1}. Therefore, we infer that similar phonon modes are responsible for the field-induced structural instability and hence the electric polarization along this direction in CsEr(MoO$_{4}$)$_{2}$.

\section{CONCLUSION}
We performed detailed experimental studies characterizing the structural, magnetic, and electric properties of the alkali rare-earth double molybdate CsEr(MoO$_{4}$)$_{2}$. We found that the room-temperature {\it P2/c} space group persists down to 0.3~K and the zero-field magnetic structure below $T_N$~$=$~0.87~K consists of antiferromagnetic chains of Er ions with Ising moments pointing along the {\it c}-axis. Most intriguingly, we identified magnetoelectric behavior with $H$ // $a$-axis below $T_N$ in a fairly narrow pocket of the $T-H$ phase diagram, consistent with a theoretical prediction \cite{02_loginov}. We attribute the strong magnetoelectric coupling in this material to a pseudo Jahn-Teller distortion that arises due to the relatively small energy gap between the Er$^{3+}$ CEF ground state doublet and first excited state. Therefore, we have shown that rare-earth elements can be solely responsible for magnetoelectric coupling through a Jahn-Teller mechanism. Our work calls for first principles calculations and experimental characterization of the electronic band structure and phonon density-of-states for CsEr(MoO$_{4}$)$_{2}$ to better understand the pJTD mechanism responsible for the magnetoelectric effect in this system.
%The magnetic properties implies that the electronic band structure of antiferromagnetic is formed by splitting the Kramers degeneracy in paramagnetic ground state due to weak exchange interaction between Er$^{3+}$ and no energy level cross occurs. We have shown that the ferroelectricity could be induced in antiferromagnetic phase. The ferroelectricity possibly originates from Jahn-Teller distortion by placing the electronic energy close to where the polar phonon density of states is maximized by external magnetic field. Therefore, multiferroicity from Jahn-Teller effect is not only possible in $3d$ transition metal compounds but also in rare earth element compounds as well. Also our study calls for detailed study of electronic band structure and phonon band structure from the first-principle calculations and group theoretical understanding to identify the exact nature of the electronic ground state and polar phonon mode responsible for ferroelectricity.

% \section*{DATA AVAILABILITY}
% Data are available from the corresponding authors upon reasonable request.

\begin{acknowledgments}
We acknowledge K.M. Taddei for providing assistance with the NPD refinements and G. Sala for performing the CsEr(MoO$_4$)$_2$ CEF point charge calculations. Q.C., Q. H. and H.D.Z. thank the support from NSF-DMR through Award DMR-1350002. This research used resources at the High Flux Isotope Reactor, a DOE Office of Science User Facility operated by the Oak Ridge National Laboratory. A portion of this work was performed at the NHMFL, which is supported by National Science Foundation Cooperative Agreement No. DMR-1157490 and the State of Florida. E. S. C. and M. L. acknowledge the support from NSF-DMR-1309146. G. H. W. and J. M. were supported by the National Natural Science Foundation of China (11774223, U1732154), and the Ministry of Science and Technology of China (2016YFA0300501). Z.W., L.L., and Z. Q. were supported by the National Natural Science Foundation of China under Grant Nos 11774352, U1832214. A portion of this work was supported by the High Magnetic Field Laboratory of Anhui Prince. 
\end{acknowledgments}

% \section*{AUTHOR CONTRIBUTIONS}
% The project was conceived by H. D. Z. The sample was synthesized by Q.C., Q. H. and H. D. Z. Dc susceptibility, ac susceptibility, high-field magnetization measurements, dielectric constant measurement and pyroelectric current measurement were performed by M. L. and E. S. C. Z. W., L. L., Z. Q., G. H. W. and J. M. carried out low temperature dc susceptibility and specific heat measurement. Neutron powder diffraction was performed by Q. C. and A. A. A. M. L., Q. C., A. A. A. and H. D. Z. analyzed the data. M. L. and A. A. A. wrote the manuscript, and all authors commented on the manuscript.

% \section*{COMPETING INTERESTS}
% The authors declare no competing financial interests.
% \\

% {\bf Corresponding authors}: Correspondence to Minseong Lee or Adam A. Aczel.


\begin{thebibliography}{50}
\bibitem{note} The b and c-axes are mislabelled in Ref.~\cite{94_khatsko} from Fig.~2 until the end of the paper. We verified this error by performing our own CEF point charge calculations for CsEr(MoO$_4$)$_2$. While our results confirm a CEF ground state doublet with predominantly $\left|J_z = \pm \frac{15}{2}\right>$ character, we find that the Ising moments point along the c-axis for this specific unit cell choice.
\end{thebibliography}
\end{document}